\documentclass[prb,a4paper,twocolumn,showpacs,floatfix,superscriptaddress,amsmath,amsfonts,amssymb,preprintnumbers,citeautoscript,aps,longbibliography]{revtex4-2}
\pdfoutput=1

\usepackage{hyperref}
\hypersetup{backref=true,pagebackref=true,hyperindex=true,colorlinks=true,breaklinks=true,
urlcolor=blue,linkcolor=blue,bookmarks=true,bookmarksopen=true,citecolor=red}

\usepackage[T1]{fontenc}\usepackage[latin1]{inputenc}

\usepackage{amsmath}
\usepackage{amssymb}
\usepackage{bbold} 
\usepackage{graphicx}
\usepackage{dcolumn,graphicx,color,booktabs,microtype,afterpage}

\usepackage{bm}
\usepackage{braket}

\usepackage{times}

\usepackage{color}
\usepackage[normalem]{ulem}
\begin{document}

\title{The interplay of  Dzyaloshinskii-Moriya and Kitaev interactions for magnonic properties of Heisenberg-Kitaev honeycomb ferromagnets}

\author{Li-Chuan~Zhang}\email[Corresponding author:~]{li.zhang@fz-juelich.de}
\affiliation{Peter Gr\"unberg Institut and Institute for Advanced Simulation, Forschungszentrum J\"ulich and JARA, 52425 J\"ulich, Germany}
\affiliation{Department of Physics, RWTH Aachen University, 52056 Aachen, Germany}
\author{Fengfeng Zhu}
\affiliation{J\"ulich Centre for Neutron Science (JCNS) at Heinz Maier-Leibnitz Zentrum (MLZ), Forschungszentrum J\"ulich, Lichtenbergstrasse 1,
D-85747 Garching, Germany}
\affiliation{Department of Physics and Astronomy, Shanghai Jiao Tong University, Shanghai 200240, China}
\author{Dongwook Go}
\affiliation{Peter Gr\"unberg Institut and Institute for Advanced Simulation, Forschungszentrum J\"ulich and JARA, 52425 J\"ulich, Germany}
\affiliation{Institute of Physics, Johannes Gutenberg University Mainz, 55099 Mainz, Germany}
\author{Fabian~R.~Lux}
\affiliation{Institute of Physics, Johannes Gutenberg University Mainz, 55099 Mainz, Germany}
\author{Flaviano Jos\'e dos Santos}
\affiliation{Peter Gr\"unberg Institut and Institute for Advanced Simulation, Forschungszentrum J\"ulich and JARA, 52425 J\"ulich, Germany}
\affiliation{Theory and Simulation of Materials (THEOS), and National Centre for Computational Design and Discovery of Novel Materials (MARVEL), \'Ecole Polytechnique F\'ed\'erale de Lausanne, 1015 Lausanne, Switzerland}
\author{Samir Lounis}
\affiliation{Peter Gr\"unberg Institut and Institute for Advanced Simulation, Forschungszentrum J\"ulich and JARA, 52425 J\"ulich, Germany}
\affiliation{Faculty of Physics, University of Duisburg-Essen, 47053 Duisburg, Germany}
\author{Yixi Su}
\affiliation{J\"ulich Centre for Neutron Science (JCNS) at Heinz Maier-Leibnitz Zentrum (MLZ), Forschungszentrum J\"ulich, Lichtenbergstrasse 1,
D-85747 Garching, Germany}
\author{Stefan~Bl\"ugel}
\affiliation{Peter Gr\"unberg Institut and Institute for Advanced Simulation, Forschungszentrum J\"ulich and JARA, 52425 J\"ulich, Germany}
\author{Yuriy~Mokrousov}\email[Corresponding author:~]{y.mokrousov@fz-juelich.de}
\affiliation{Peter Gr\"unberg Institut and Institute for Advanced Simulation, Forschungszentrum J\"ulich and JARA, 52425 J\"ulich, Germany}
\affiliation{Institute of Physics, Johannes Gutenberg University Mainz, 55099 Mainz, Germany}

\begin{abstract}
\noindent 
The properties of Kitaev materials are attracting  ever increasing attention owing to their exotic properties. In realistic two-dimensional materials, Kitaev interaction is often accompanied by the Dzyloshinskii-Moriya interaction, which  poses a challenge of distinguishing their magnitude separately. In this work, we demonstrate that it can be done by accessing magnonic transport properties. 
By studying honeycomb ferromagnets exhibiting Dzyaloshinskii-Moriya and Kitaev interactions simultaneously, we reveal  non-trivial  magnonic topological properties accompanied by intricate magnonic transport characteristics as given by  thermal Hall and magnon Nernst effects. We also investigate the effect of  a  magnetic field, showing that it does not only break the symmetry of  the system  but also  brings drastic modifications to magnonic topological transport properties, which serve as hallmarks of the relative strength of anisotropic exchange interactions.  Based on our findings, we suggest strategies to estimate the importance of Kitaev interactions  in real materials. 
\end{abstract}

\maketitle


{\it Introduction.} Recently, layered magnetic materials with highly-anisotropic Kitaev-like spin interactions originated in spin-orbit coupling (SOC) are attracting increasing attention~\cite{chaloupka2010kitaev,chaloupka2013zigzag,janssen2016honeycomb,banerjee2016proximate,kasahara2018majorana}. 
The realization of the celebrated Heisenberg-Kitaev
model has been to date verified in layered irridates A$_2$IrO$_3$~(A= Li; Na)~\cite{singh2012relevance,choi2012spin,ye2012direct, chun2015direct}, $\alpha$-RuI$_3$~\cite{sears2015magnetic,banerjee2016proximate} and CrI$_3$~\cite{lee2020fundamental}. It is known that depending on specific parameters, the  Heisenberg-Kitaev model can host gapless or gaped spin liquid states~\cite{banerjee2016proximate,kasahara2018majorana}, 
and that a topologically ordered phase can be achieved by applying an external magnetic field~\cite{kitaev2006anyons, jiang2011possible}. This indicates that the Heisenberg-Kitaev model hosts a rich phase diagram~\cite{janssen2016honeycomb,mcclarty2018topological}, and as such, Kitaev materials present a promising material platform for realization of novel applications in the areas of  topological quantum computing and spintronics~\cite{nayak2008non,vzutic2004spintronics}.

One of the most natural ways to extract the properties of Kitaev interaction lies in the analysis of the magnon spectra of a given Kitaev material, which naturally incorporates the effect of exchange interactions as well as of the magnetic field~\cite{wu2018field,janvsa2018observation,cenker2020direct}. However, it has been recently shown that while the Kitaev interaction can realize topological magnon bands in honeycomb ferromagnets~\cite{mcclarty2018topological,joshi2018topological}, its effect can be generally similar to that of the SOC-driven Dzyaloshinskii-Moriya interaction~(DMI)~\cite{dzyaloshinsky1958thermodynamic,Moriya60}. In fact, the second-nearest-neighbor DMI, which is allowed by symmetry in  honeycomb  materials ~\cite{kim2016realization,chen2018topological},  has been estimated explicitly from  ab-initio calculations of monolayer CrI$_3$~\cite{xu2020topological}. Therefore, the same magnon dispersion can be interpreted based either on DMI, Kitaev interaction, or their combination. 
In order to distinguish whether the system is DMI or Kitaev interaction dominated, magnonic properties other than the dispersion have to be investigated in detail. 

\begin{figure}[ht!]
\includegraphics[width=\linewidth]{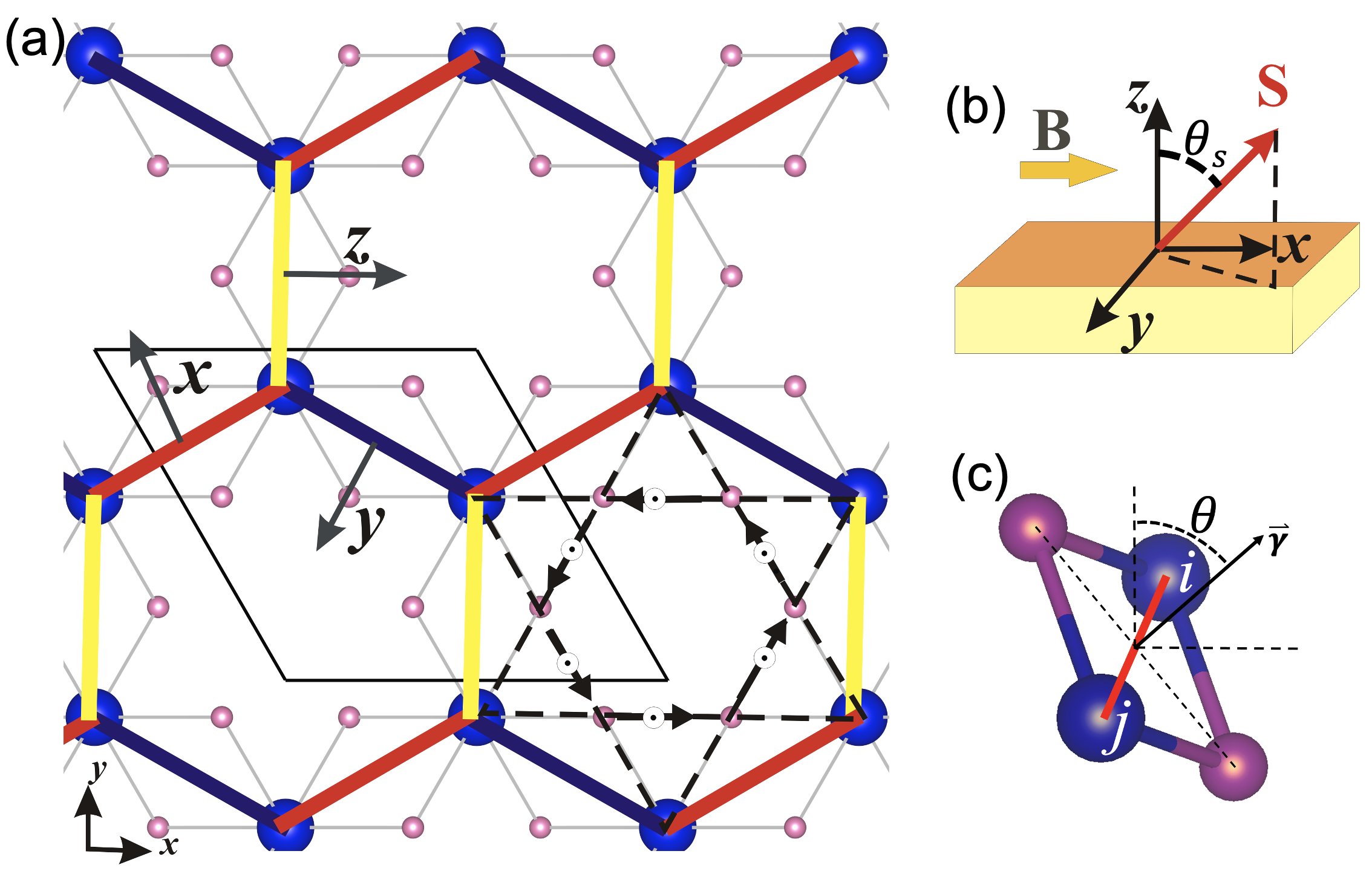}
\caption{(a)~Sketch of the structure of honeycomb CrI$_3$ monolayer. The unit cell is outlined with a thin black line, where blue balls represent  Cr$^{+3}$ ions and pink balls are iodide ions. The  Kitaev bonds $\textbf{\emph x}$ (red), $\textbf{\emph y}$ (dark blue), $\textbf{\emph z}$ (yellow) are indicated with thick colored lines.
The arrows mark the second-nearest-neighbor bond orientations along black dotted lines that share a common sign of out-of-plane DM vector. (b)~Schematic diagram of the influence of an in-plane magnetic field $
\mathbf B$ on the spin direction  $\mathbf S$ whose polar angle is defined as $\theta_s$.  
(c)~The perspective view of Cr$_2$I$_2$ plane, where its normal vector is marked as $\hat{\gamma}$ and  the Kitaev angle is defined as the polar angle of $\hat {\gamma}$. }
\label{structure}
\end{figure}

In this work, we investigate the magnonic properties of  honeycomb ferromagnets exhibiting Kitaev and DMI interactions exposed to a magnetic field, see Fig.~1. Based on the topological analysis of the magnonic states, we characterize the spectra of the model and make predictions concerning the behavior of the thermal Hall and magnon Nernst conductivity in response to changes in the paramters of the model, proposing a  strategy to distinguish whether the system is dominated by Kitaev interaction or DMI.

\begin{figure*}[ht!]
    \centering\vspace{-1pt}
    \includegraphics[width=0.95\linewidth]{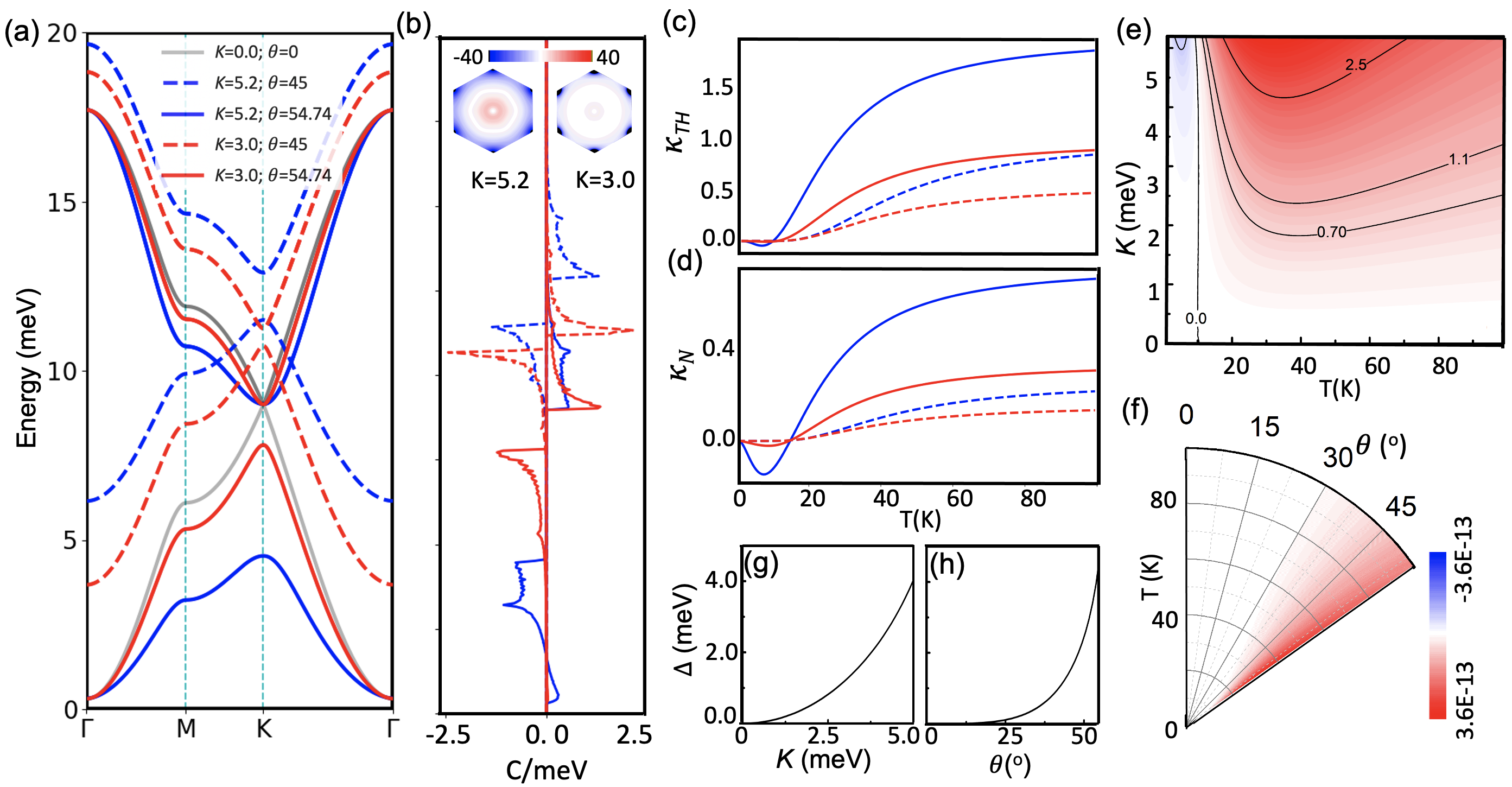}
    \caption{ (a)~The comparison of magnon dispersions along  high-symmetry lines for different values of the Kitaev parameters. Grey, blue and  red lines correspond to the $K$ values of 0, 5.2 and 3\,meV.  The dashed and solid lines correspond to the Kitaev angles $\theta$ of 45$^\circ$ and 54.74$^\circ$, respectively.  The  corresponding energy-resolved Chern number is shown in (b).  The Berry curvature distribution of the first magnon branch in the first Brillouin zone for different $K$-values with $\theta=54.74^{\circ}$ is shown in the inset of (b). The color map ranges from $-$40 to 40\, in arb. units, and the exceeding values are marked with black.  The temperature dependence of thermal Hall conductivity $\kappa^{xy}_\mathrm{TH}$ and magnon Nernst conductivity $\kappa^{xy}_\mathrm{N}$ is shown in (c) and (d), in units of $10^{-11}$\,W/K and $k_B/2\pi$, respecitively. (e)~The map of  $\kappa^{xy}_\mathrm{TH}/T$ as a function of temperature $T$ and parameter $K$ at the Kitaev angle of $\theta=54.74^{\circ}$. (f)~The map  of $\kappa^{xy}_\mathrm{TH}/T$ as a function of $T$ and $\theta$ at the constant $K$-value of 5.2\,meV. The units of the color maps in (e)-(f) are chosen as W/K$^2$. The magnon band gap $\triangle$ as a function of  $K$ (at $\theta$=$54.74^{\circ}$) and  $\theta$ (at $K$=5.2 meV) is shown in (g) and (h), respectively.  }
    \label{Without_D_B}
\end{figure*}

{\it The model and  the method}. We consider the effective spin Hamiltonian on a two-dimensional ferromagnetic honeycomb lattice,  sketched in Fig.~1, given by
\begin{equation}
\begin{split}
    H=&-\sum_{i,j}J_{ij}  \mathbf{S}_i \cdot \mathbf{S}_j -K\sum_{\braket{ij}^{\gamma}} S_i^{\gamma}S_j^{\gamma}-\sum_{ij}\mathbf{D}_{ij} \cdot (\mathbf{S}_i\times \mathbf{S}_j)\\
     &-\sum_iA(\hat{n}_i\cdot \mathbf S_i)^2
     - \mathbf{B}\cdot\mu_Bg\sum_i \mathbf{S}_i,\, 
\end{split}
\label{Ham}
\end{equation}
where the $J_{ij}$ coefficients mediate the isotropic Heisenberg exchange interaction between spins $\mathbf S_i$ and $\mathbf S_j$ on sites $i$ and $j$, and the second term is due to the anisotropic Kitaev interaction, where $S_i^{\gamma}$= $\mathbf S_i\cdot \hat{\gamma}_{ij}$ with $\hat{\gamma}_{ij}$ being the Kitaev vector determined by the sites $i$ and $j$. The second-nearest-neighbor DMI is represented by the third term with DMI vectors $\mathbf{D}_{ij}$ pointing out of plane, as required by the symmetry of the structure,~i.e.,~$\mathbf{D}_{ij}=(0, 0, D_{ij}^z)$.  Additionally, we add a single-ion anisotropy term  with respect to the local  easy axis $\hat{n_i}$ (choosing it to be the unit vector along the $z$-direction), and the energy of Zeeman coupling to the magnetic field $\mathbf{B}$, with $\mu_B$ as Bohr magneton and  $g$-factor of 2.

The structure that we consider here, Fig.~\ref{structure}, is a representative of the Kitaev materials such as 
$\alpha$-RuI$_3$~\cite{sears2015magnetic,banerjee2016proximate} and CrI$_3$~\cite{lee2020fundamental}, and for simplicity we refer to our studied system as CrI$_3$ in the following. 
Referring to experimental data on the latter material~\cite{lee2020fundamental}, we set approximate values for the nearest-neighbor exchange interaction $J = 0.2$\,meV, the Kitaev interaction $K = 5.2$\,meV, and the spin moment magnitude $S = 1.5$.
We consider an easy-axis anisotropy energy of $A=0.1$\, meV  chosen so as to ensure that the ground state is ferromagnetic along the $z$-axis.
As displayed in Fig.~\ref{structure}(c), $\hat{\gamma}_{ij}$ is defined as the normal vector to the  Cr$_2$I$_2$ plane spanned by  Cr ions $i$ and $j$, and the nearby I atoms.  Respectively, the Kitaev vector corresponding to the yellow bond in Fig.~\ref{structure}(a) and marked with $ {\textbf{\emph z}}$ is chosen as $\hat{\gamma}_{\textbf{\emph z}}$=(sin$\theta$, 0 cos$\theta$)=($\frac{\sqrt{2}}{\sqrt{3}}$, 0, $\frac{1}{\sqrt{3}})$~\cite{lee2020fundamental,aguilera2020topological}, where the Kitaev angle $\theta$ is about 54.74$^{\circ}$ for the case of CrI$_3$. The Kitaev vectors for red and blue bonds are determined analogously. 
More details concerning the model can be found in the Supplementary Material.

Further, the Holstein-Primakoff transformation~\cite{holstein1940field} is employed  to rewrite the Hamiltonian in terms of bosonic ladder operators $a_i$ and $a_i^\dagger$.  In the transformed spin-wave Hamiltonian, we keep only the quadratic terms in the spin operators  and a Fourier transform of the bosonic ladder operators is performed to rewrite the problem in the  momentum space.  
The fourier-transformed Hamiltonian, denoted as $H_2$, thus becomes a $2n\times 2n$ matrix, where $n=2$ stands for two  atoms in the unit cell of honeycomb lattice~\cite{SantosSantosDias18,zhang2020magnonic}. 
We diagonalize  the dynamical matrix of $H_2$  based on the commutation relation: $
    \mathrm{i}\frac{\mathrm{d}\Phi(\mathbf k)}{\mathrm{d}t}=[\Phi(\mathbf k),H_2(\mathbf k)]=D\Phi(\mathbf k),
$
where the dynamical matrix is given by $D=\hat{g}H_2$ with $\hat{g}=[(\mathbb{1}, 0), (0, -\mathbb{1})]$,  $\mathbb{1}$ as the $n \times n$ identity matrix, and a basis is chosen as $\Phi({\mathbf k})=[a_{1k}, ~a_{2k},~ a_{1-k}^{\dag}, ~a_{2-k}^{\dag}]^T$.
Only positive real eigenvalues of the  dynamical matrix $D$ are considered and the stability of the system is confirmed  when there are two non-negative eigenvalues for each vector $\mathbf k$.
We employ the magnon Berry curvature formalism to investigate the topological properties of the model, with the magnon Berry curvature of the $n$th spin-wave branch $\Omega_{n\mathbf k}^{xy}$ evaluated according to:
\begin{equation}
 \Omega_{n\mathbf k}^{xy}=-2\,\mathrm{Im}\sum_{m\neq n}
 \frac{\braket{\Psi_{n\mathbf{k}}|\frac{\partial D(\mathbf{k})}{\partial k_x}|\Psi_{m\mathbf{k}}}\braket{\Psi_{m\mathbf{k}}|\frac{\partial D(\mathbf{k})}{\partial k_y}|\Psi_{n\mathbf{k}}}}{(\epsilon_{n\mathbf{k}}-\epsilon_{m\mathbf{k}})^2} \, ,
 \label{curvature}
 \end{equation}
where $|\Psi_{n\mathbf k}\rangle$ is the right eigenstate of the spin-wave Hamiltonian with the energy $\epsilon_{n\mathbf k}$.

\begin{figure*}[ht!]
    \centering\vspace{-1pt}
    \includegraphics[width=0.95\linewidth]{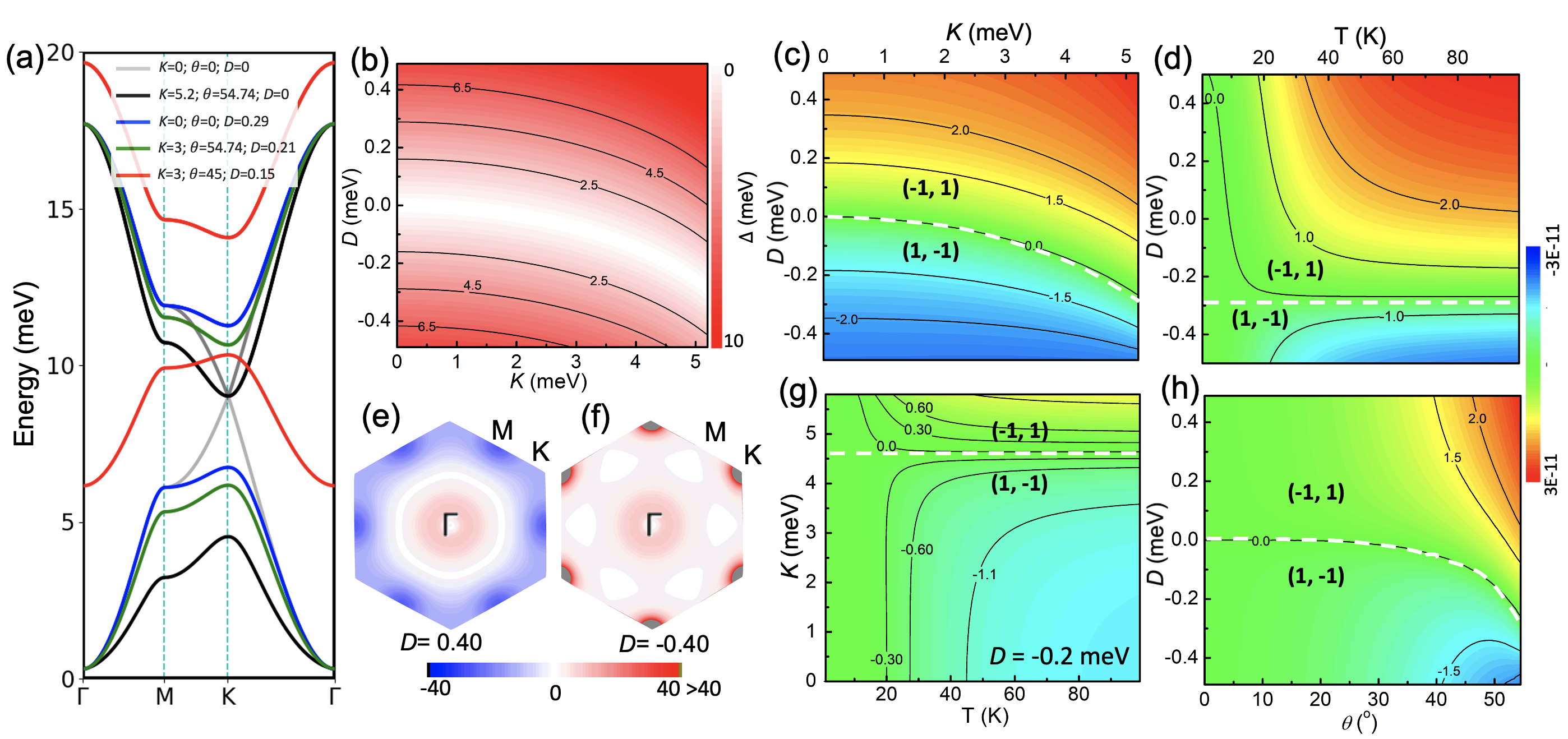}
    \caption{ (a)~The comparison of magnon dispersions  with different Kitaev and DMI parameters specified in the legend (in the units of meV and degrees).  The magnon dispersions represented with black, blue, green and red lines have almost the same band gap $\triangle$ between the two branches.  (b)~Band gap $\triangle$ as a function of $D$ and $K$ at $\theta=54.74^{\circ}$. (c) Dependence of thermal Hall conductivity $\kappa^{xy}_\mathrm{TH}$ on $K$ and  $D$ at $T=100$\,K and $\theta=54.74^{\circ}$. The $T$-dependence of thermal Hall conductivity $\kappa^{xy}_\mathrm{TH}$ is shown in (d-g). (d)  $\kappa^{xy}_\mathrm{TH}$ as a function of $T$ and  $D$, with $K$ and $\theta$ corresponding to the case of CrI$_3$. (e-f)~The corresponding BZ distribution of the Berry curvature  (in arb. units)  for the first band  with  $D=+0.4$\,meV (e) and $-$0.4\,meV (f) and the Kitaev interaction is the same of (d).  (g)~  $\kappa^{xy}_\mathrm{TH}$ as a function of $T$ and $K$ for  $\theta=54.74^{\circ}$ and $D$=$-$0.2\,meV. (h)   $\kappa^{xy}_\mathrm{TH}$ as a function of $\theta$ and  $D$ at $T=100$\,K and $K=5.2$\,meV.  The corresponding color map is in the units of W/K. 
 }
    \label{D_effect}
\end{figure*}

The topological thermal Hall effect of magnons is the generation of a transverse thermal Hall voltage under an applied longitudinal temperature gradient~\cite{OnoseIdeue10, HirschbergerChisnell15}. Based on the expression for the Berry curvature, the transverse thermal Hall conductivity $\kappa_\mathrm{TH}^{xy}$ of the system~\cite{mook2014magnon} is calculated according to the following expression:
\begin{equation}
 \kappa^{xy}_\mathrm{TH}=-\frac{k_\mathrm{B}^2 T}{(2\pi)^2 \hbar} \sum_n\int_{\rm BZ} c_2 (n_B(\epsilon_{n\mathbf k}))\, \Omega^{xy}_{n\mathbf{k}}\,d\mathbf{k} \, ,
 \label{equ2}
\end{equation}
where $c_2(\tau)=(1+\tau)\ln^2[(1+\tau)/\tau]-\ln^2 \tau-2\mathrm{Li}_2 (-\tau)$,
 $\mathrm{Li}_2$ represents the dilogarithm function,  $n_B(\epsilon_{n\mathbf k})=(e^{\epsilon_{n\mathbf{k}}/k_\text{B}T}-1)^{-1}$ is the Bose-Einstein distribution function, and BZ denotes the Brillouin zone.
In addition, the magnon Nernst conductivity, $\kappa_\mathrm{N}^{xy}$, which  represents  the magnon-mediated transverse transport of spin, is calculated  based on the following expression:
\begin{equation}
\begin{split}
    \kappa_\mathrm{N}^{xy}=-\frac{k_\mathrm{B}}{(2\pi)^2 }\sum_n\int_{\rm BZ} c_1(n_\mathrm{B}(\epsilon_{n\mathbf k}))\,\Omega_{n\mathbf k}^{xy}
     \, d\mathbf k \, ,
\end{split}
\label{eq:tone}
\end{equation}
with $c_1(\tau)=\int_0^{\tau}\ln[(1+t)/t]dt=(1+\tau)\ln(1+\tau)-\tau\ln \tau$~\cite{kovalev2016spin,zyuzin2016magnon}.

{\it The Heisenberg-Kitaev model.} We first ignore the effect of the magnetic field and DMI,  and focus on  the magnonic transport properties of the simplified  Heisenberg-Kitaev model. 
As the Kitaev angle $\theta$ can be different in different Kitaev materials~\cite{xu2018interplay,lee2020fundamental}, we investigate the Heisenberg-Kitaev model by varying the $\theta$ angle  and the magnitude of $K$, assuming that the sign of the latter remains positive. We further keep the value of $J+K/3$ constant so as to ensure that the ground state has the same energy.  By comparing the band dispersions for different $K$ and $\theta$ values shown in Fig.~\ref{Without_D_B}(a), we find that the band gap   between the two modes $\triangle$ is enlarged as either $K$ or $\theta$ increases, see Fig.~2(g,h).   
Meanwhile,
a larger $K$ not only decreases the spin stiffness at the $\Gamma$ point but also opens a larger  band gap at $\mathbf K$, as shown in Fig.~\ref{Without_D_B}(a, h).
As a larger $\theta$ enhances the effect of anisotropic exchange interaction, the single-ion anistropy energy is introduced to  ensure the stability of the system.

We address the topological character of the magnonic bands by computing the Chern number $C_n$, given by $C_n=\frac{1}{2\pi}\int\Omega_{n\mathbf k}^{xy} \,d\mathbf k$, where the integral is performed over the first BZ, and $n$ is the $n$th magnon branch. The calculated Chern numbers are $-$1 and $+$1 for the first and second branches in Heisenberg-Kitaev model. As shown in Fig.~\ref{Without_D_B}(b), 
the energy-dependent Chern number defined as an integral of the Berry curvature at a given energy, as well as the Berry curvature distribution of the first branch indicate that the largest contributions to the Chern number come from  around the $\mathbf K$-point. Besides, the observed Chern number variation and Berry curvature distribution are quite non-trivial in energy and in the reciprocal space,  which brings about the unusual topological transport properties as manifested in the unusual temperature dependence of the thermal Hall conductivity and the magnon Nernst conductivity. 

As shown in Fig.~\ref{Without_D_B}(c-d),  a sign change with increasing temperature $T$ of $\kappa^{xy}_\mathrm{TH}$ and  $\kappa^{xy}_\mathrm{N}$ is clearly obtained for the values of $K=5.2$\,meV and $\theta=54.74^{\circ}$, which is in line with the observations for Kitaev materials~\cite{kasahara2018unusual,hentrich2019large}. This can be explained by a variation in the sign of the energy-dependent Chern number in the energy region of $1-2$\,meV for these specific values of $K$ and $\theta$, which is absent for smaller values of Kitaev parameters. 
For smaller $K$ and $\theta$, the Berry curvature magnitude rises at much higher energies, which  explains the overall suppression of thermal Hall and magnon Nernst conductivities that we observe.
To emphasize this effect further, we plot the dependence of $\kappa^{xy}_\mathrm{TH}/T$ on temperature and parameters $K$ and $\theta$ separately in Fig.~\ref{Without_D_B}(e-f). In this figure, we observe that regardless of the sign of $\kappa^{xy}_{\mathrm{TH}}/T$, 
its absolute value always increases with  K  and  $\theta$ at a given temperature. Similar conclusions can be drawn for the magnon Nerst conductivity. 
In Fig.~\ref{Without_D_B}(f), the range of considered angles is limited by $54.75^{\circ}$ owing to the fact that the system becomes unstable if the single-ion anisotropy energy remains unchanged. More details concerning the $\theta$-dependence of the  magnon Nernst effect are provided in the Supplementary Material.

\begin{figure*}[t!]
    \includegraphics[width=0.7\linewidth]{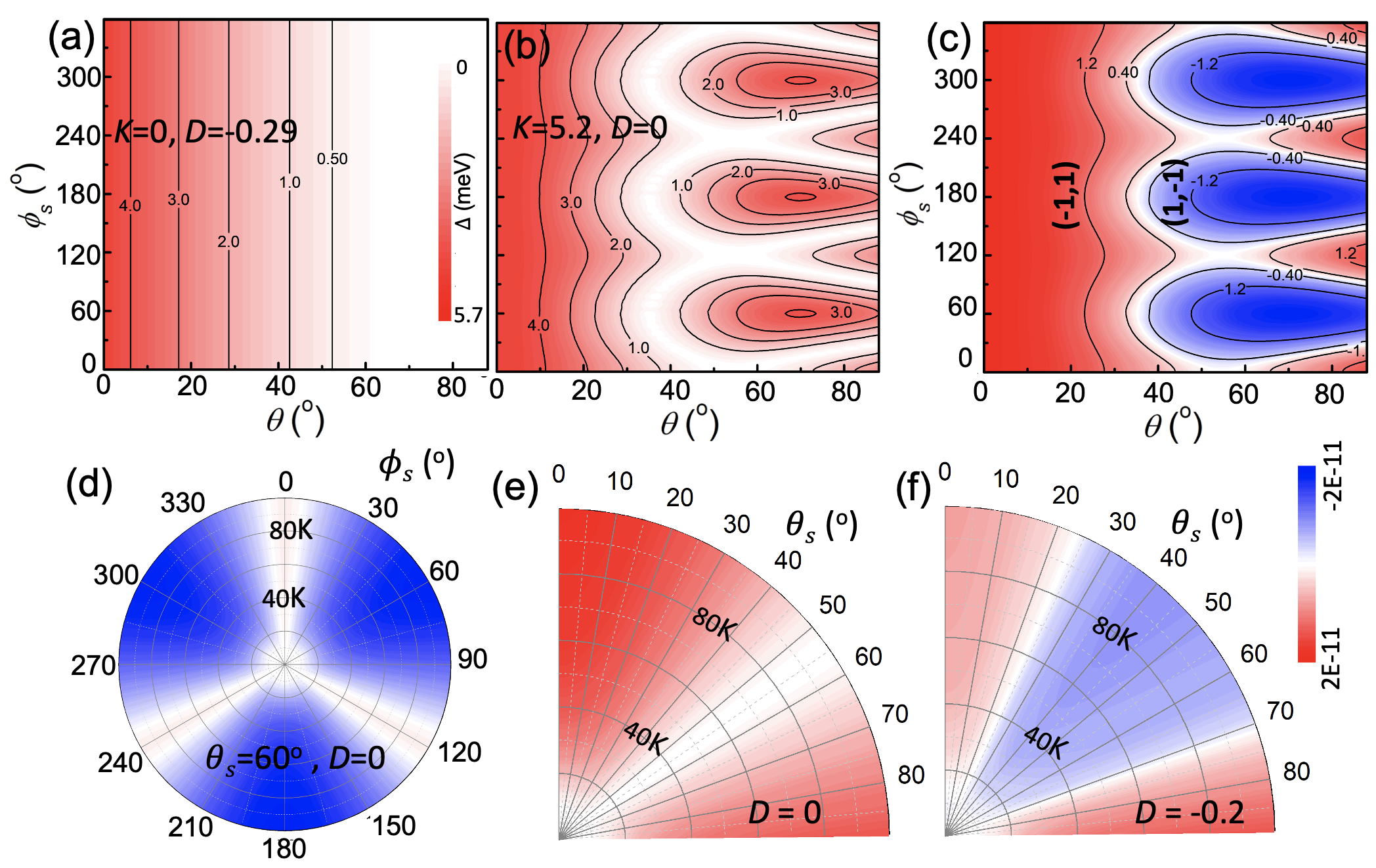}
    \caption{  (a,b)~Evolution of the band gap $\triangle$  with the angles $\theta_s$ and  $\phi_s$ in the Heisenberg-DMI model ($K=0$) (a), and Heisenberg-Kitaev model ($D=0$) (b). (a, b) share the same color map and the units are meV. (c)~Thermal Hall conductivity $\kappa^{xy}_{\mathrm{TH}}$ as a function of $\theta_s$ and $\phi_s$ at $T=100$K (same parameters as in (b)). (d)~  $\kappa^{xy}_{\mathrm{TH}}$ as a function of temperature and  $\phi_s$ at $\theta_s=60^{\circ}$
    (e-f)  $\kappa^{xy}_{\mathrm{TH}}$ as a function of temperature and $\theta_s$ for different values of $D$ assuming  $\phi_s=0^{\circ}$ and $K=5.2$\,meV.  (c-f) share the same color map in units of W/T.}
    \label{rotate_B}
\end{figure*} 

{\it Impact of DMI.} Next,  we investigate the impact of DMI on the magnonic transport properties of the  Heisenberg-Kitaev model. 
As shown in Fig.~\ref{D_effect}(a), our results indicate that both DMI and Kitaev interaction can modify the magnon dispersion and open a gap at the crossing point at $\mathbf K$. The difference in the impact of  DMI and Kitaev interactions is that  the latter strongly influences the  shape of magnon dispersion, whereas the DMI mainly influences the dispersion around the $\mathbf K$ point. As also visible in Fig.~\ref{D_effect}(a-b), a band gap of the same magnitude $\triangle$ can be realised by a combination of different DMI and Kitaev parameters. 
For instance, the experimental  magnon dispersion of CrI$_3$ can be fitted well with both Heisenberg-Kitaev model or Heisenberg-DMI model~\cite{chen2018topological,lee2020fundamental}. In this context, the relevance of a given model can be probed by accessing its topological transport properties and comparing them to experiments.

The topological thermal Hall conductivity $\kappa^{xy}_\mathrm{TH}$ modulated by Kitaev parameters ($\theta, K$) and DMI ($D$) is shown in Fig.~\ref{D_effect}(c-d, g-h). The topological phase boundary marking different sets of  ($C_1$,$C_2$) Chern numbers is shown with a white dashed line.  As the sign of the Berry curvature generally changes in the BZ for the first branch, Fig.~\ref{D_effect}(e-f), the zero isoline of $\kappa^{xy}_\mathrm{TH}$ does not generally coincide with the phase boundary, which is different from the purely DMI-mediated system~\cite{mook2014magnon,zhang2020magnonic}. Similar to Fig.~\ref{Without_D_B}(e), the sign change of   $\kappa^{xy}_\mathrm{TH}$ is observed with increasing $T$ in Fig.~\ref{D_effect}(d-g) in the topological phase marked as ($-$1,$+$1).   This feature can be explained by the fact that while DMI mainly influences the magnonic states around $\mathbf K$ or $\mathbf M$ points, the Berry curvature around $\mathbf {\Gamma}$ point is mainly determined by the Kitaev interaction, see Fig.~\ref{D_effect}(e-f) and insets of Fig.~\ref{Without_D_B}(b). The total contribution to $\kappa^{xy}_\mathrm{TH}$ thus presents a subtle competition between Berry curvature contributions from around these points, whose overall sign depends on the interplay between the parameters.
The phase diagrams of $\kappa^{xy}_\mathrm{TH}$ with respect to $\theta$, $K$ and $D$ at $T=100$\,K are shown in Figs.~\ref{D_effect}(c, h). Consistent with the discussion above, the magnitude of  $\kappa^{xy}_\mathrm{TH}$  is directly determined by the strength of $K$ and magnitude of $\theta$.
Notably, at a given $K$, the sign of $\kappa^{xy}_\mathrm{TH}$ can be adjusted by the sense of DMI.
Similar observations can be made also for the magnon Nernst conductivity, as shown in the Supplementary Material.

{\it The effect of a magnetic field}. Finally, we explore the effect of an external in-plane magnetic field,~Fig.~\ref{structure}(b). As a result of the magnetic field $\mathbf B=\textup{B}(\cos\phi_s, \sin\phi_s, 0)$ the spins of the Kitaev magnet
are inclined into the plane:
 ${\mathbf{S}}_i = \textup{S} (\sin\theta_s\cos\phi_s, \sin\theta_s\sin\phi_s, \cos\theta_s)$,
 where $\theta_s$ and $\phi_s$ represent the polar angle and azimuthal angle respectively. 
The relationship between the strength of the field $B$ and the inclination angle $\theta_s$ is given by $\sin\theta_s=g\mu_B\textup{B}/2A\textup{S}$ (see more details in Supplementary Material). 
As shown in Fig.~\ref{rotate_B}(a-b) and Supplementary Material,  both polar angle $\theta_s$ and azimuthal angle $\phi_s$ have an influence on the magnon dispersion. Especially $\phi_s$  has a strong impact on  the  band gap when $\theta$ is larger than $40^{\circ}$ at finite $K$, while the band gap is only influenced by the  polar angle $\theta_s$ in the  Heisenberg-DMI model ($K=0$).  Furthermore, as we show in the Supplementary Material,
the  $C_3$ symmetry  of the magnon dispersion is broken if the polar angle is nonzero assuming non-vanishing Kitaev interaction.

We draw the topological phase diagram of thermal Hall conductivity as a function of $\theta_s$ and $\phi_s$ in Fig.~\ref{rotate_B}(c). When $\theta_s$ is smaller than $40^{\circ}$, the system resides in the ($-$1, $+$1) phase, and the influence of $\phi_s$ is suppressed. However,  $\kappa^{xy}_\mathrm{TH}$
exhibits a very non-trivial dependence on $\phi_s$ when the system enters the ($+$1, $-$1) phase upon increasing $\theta_s$. 
The strong dependence of $\kappa^{xy}_\mathrm{TH}$
on $\phi_s$ and $\theta_s$ is also visible in the temperature-dependence plots shown in Fig.~\ref{rotate_B}(d, e, f).
As becomes apparent from Fig.~\ref{rotate_B}(d), the $C_3$ symmetry of the conductivity is preserved, in line with the  symmetry of the Kitaev interaction on a honeycomb lattice. Moreover, from Figs.~\ref{rotate_B}(e,f)
we observe a strong influence of the DMI  on the magnitude and angular dependence of the thermal Hall conductivity. Overall our result reveal a rich landscape of thermal Hall effect of Kitaev ferromagnets exposed to an external magnetic field.

{\it Discussion}. In our study we investigate the magnonic properties of honeycomb feromagnets with DMI and Kitaev interaction subject to an external magnetic field. On the one hand, we observe intricate magnonic transport characteristics, which have been observed in Kitaev materials~\cite{kasahara2018unusual,hentrich2019large} that we attribute to the non-trivial Berry phase properties of the system. On the other hand,  our results demonstrate a rich magnonic topological phase diagram drawn as a function of Kitaev parameters, DMI and magnetic field strength.  Since the magnitude of the latter effects can be adjusted through~e.g.~application of strain~\cite{xu2020possible} or electric field~\cite{koyama2018electric}, our investigation provides a good reference point for designing the magnonic properties of  candidate Kitaev materials.
Our findings bare significant  relevance  given that although several Kitaev materials have been discovered to date (e.g.~\cite{singh2012relevance,choi2012spin,ye2012direct, chun2015direct,sears2015magnetic,banerjee2016proximate}), it is still not clear how to judge the relative importance of Kitaev interaction with respect to DMI. 

From the perspective of magnons, based on the results of our work, we  propose a strategy to disentangle the two types of interactions from each other: 
if an application of an external in-plane magnetic field
brings along a significant modification of the shape of the magnon dispersion and a strong variation of the magnonic properties as a function of the in-plane direction of the field, then the system is dominated by Kitaev interaction rather than DMI. Additionally, the changes of sign in the thermal transverse characteristics as  a function temperature or strength of an external magnetic field can serve as another indication of the prominence of the Kitaev interaction in the system. These simple criteria can potentially enable a magnonic characterization of exchange interactions of Kitaev materials, and pave the way to employing magnonic topology for designing their exotic properties.

{\it Acknowledgements}.
This project was supported by the China Scholarship Council (CSC) (Grant No.~[2016]3100).  
The work was also supported by the Deutsche Forschungsgemeinschaft (DFG, German Research Foundation) $-$ TRR 173 $-$ 268565370 (project A11), TRR 288 $-$ 422213477 (project B06). We  acknowledge  funding  under SPP 2137 ``Skyrmionics" of the DFG.
 We gratefully acknowledge financial support from the European Research Council (ERC) under the European Union's Horizon 2020 research and innovation program (Grant No. 856538, project "3D MAGiC"). We gratefully acknowledge computing time on the supercomputers of J\"ulich Supercomputing Center, and at the JARA-HPC cluster of RWTH Aachen.

\bibliographystyle{my-apsrev}
\bibliography{Kitaev}\vspace{-3pt}

\end{document}